\documentstyle[12pt]{article}

\newcommand {\no} {\nonumber}

\def\beq{\begin{eqnarray}}
\def\eeq{\end{eqnarray}}

\def\gl#1{(\ref{#1})}
\def\L#1{{\cal L}_{#1}}

\begin{document}

\thispagestyle{empty}
\begin{titlepage}

\begin{flushright}
JINR preprint E2-96-119 \\
April 1996 \\
\end{flushright}

\vspace{3cm}
\renewcommand{\thefootnote}{\fnsymbol{footnote}}
\begin{center}
\Large \bf
Kaon polarizability in the Nambu-Jona-Lasinio model \\
at zero and finite temperature
 \\
\end{center}
\vspace{0.5cm}
\begin{center}
D.\ Ebert\footnotemark[1]{}
\footnotemark[2],
M.\ K.\ Volkov\footnotemark[1]\\
\sl {Joint Institute for Nuclear Research, Dubna, Russia.}\\
\end{center}
\vspace{0.6cm}
\begin{abstract}
\noindent
\large{
Using recent data for the decays
$f_0 \to \gamma\gamma$,
$f_0 \to \pi \pi$,
we determine
the mixing angle of scalar mesons
in a chiral quark $\sigma$-model.
This value is employed to
analyze of the kaon polarizability.
It is shown that pole diagrams from
intermediate scalar mesons and their
mixing angle
significantly affect the electromagnetic
polarizability of charged and neutral kaons.
Our results are compared with
other models and
the results of the chiral symmetry limit.
The temperature dependence of the kaon polarizabilities
is investigated.}
\end{abstract}
\vspace{0.3cm}
\setcounter{footnote}{1}
\footnotetext{Supported by
{\it Deutsche Forschungsgemeinschaft} under contract 436 RUS 113/29.}
\setcounter{footnote}{2}
\footnotetext{On leave of absence of
 Institut f\"ur Physik, Humboldt--Universit\"at zu Berlin,\\
 Invalidenstra\ss e 110, D--10115 Berlin, Germany}
\vfill
\end{titlepage}

\renewcommand{\thefootnote}{\arabic{footnote}}
\setcounter{footnote}{0}
\setcounter{page}{1}

\begin{center}
\Large \bf
Каонная поляризуемость в модели Намбу-Иона-Лазинио \\
при нулевой и конечной температуре \\
\vspace{0.5cm}
Д. Эберт, М.К. Волков \\
\vspace{0.6cm}

\large{Аннотация}
\end{center}
\begin{quote}
\large{
 Используя современные экспериментальные данные по раcпадам
$f_0 \to \gamma\gamma$, $f_0 \to \pi \pi$,
мы определяем угол смешивания скалярных мезонов в киральной кварковой
$\sigma$-модели. Это значение используется для вычисления каонной
поляризуемости. Показано, что скалярные полюсные диаграммы и
угол смешивания скалярных мезонов играют важную роль в определении
значения поляризуемости заряженных и нейтральных каонов.
Наши результаты сравниваются с результатами других моделей и
случаем кирально-симметричного
предела. Определяется температурная зависимость каонной
поляризуемости.}
\end{quote}

\newpage

\section{Introduction}

There is a renewed interest in electromagnetic
polarizabilities of the pion and kaon which
together with other low-energy parameters like
electromagnetic radii
provide us with useful information on the internal structure of
mesons (for a review on recent results obtained in
different models see e.g.\ refs.~\cite{1,2,3}).
Some of us calculated these polarizabilities
earlier within nonlinear chiral hadron theories \cite{4,5}
as well as on the basis of a bosonized NJL-model
\cite{8} leading to a linear $\sigma$-model
\cite{6,7}.
In the linear $\sigma$-model pole diagrams arising from
intermediate scalar mesons\footnote{
The $\epsilon$ meson
is considered to be a broad resonance with mass
and width which are not presently
completely fixed yet \cite{9}.
We shall use here the theoretical value of the mass
$M_\epsilon^{theor} = 650$~MeV and the new
experimental value $M_\epsilon^{exp} = 750$~MeV \cite{9}.
The masses of the other scalar mesons are given by
$M_{f_0} = 980$~MeV and $M_{a_0} = 982$~MeV \cite{10}.}
$\epsilon$, $f_0(980)$ and $a_0(980)$ turn out to play an important
role in the calculation of meson polarizabilities. In particular, for
describing the physical isoscalar scalar mesons $\epsilon$ and
$f_0$ within an underlying quark model, one has to take into
account the mixing angle $\theta$ of the $(u,d)$ and $(s)$ quark
content of these mesons (see \cite{6}).

In the latter paper strong and radiative decays of
scalar mesons were investigated solely on the level of
quark loop triangle diagrams. In the papers
\cite{Vainshtein,Scadron}
it was however argued that additional
meson loop contributions can
play an important role for the description
of the scalar meson decays.\footnote{Note
that in the case of $a_0 \rightarrow \gamma\gamma$
the additional effect of meson loops only weakly changes
the result in contrast with the result of ref. \cite{Scadron}
where the effect of 50\% was found, and quark and meson
loops interfere in a destructive way, while we found
both effects to be additive. In \cite{Vainshtein} only
scalar meson loops were considered.}
This idea does not contradict the usual $1/N_c$ expansion
since both quark and meson loops turn out to be of the
same order in $N_c$ for this case. Stimulated by these
arguments we have recalculated the decay amplitudes of
scalar mesons.
We found that pion and espacially kaon loops
turned out to be indeed essential for the description
of the decay $f_0 \to \gamma\gamma$, which is crucial
for the determination of the mixing angle
on the basis of new experimental data.
We find a value $\theta=23^\circ$ for the mixing
angle, which would change to $\theta=-18^\circ$ if
meson loops were omitted.

Clearly, changes of the mixing angle also significantly
influence other physical quantities, like the
kaon polarizability. Therefore,
we apply the new results to estimate
the kaon polarizability.

In the last part of our paper, we have investigated
the temperature dependence of the kaon polarizabilities.
The behaviour of hadrons at finite $T$, espacially
near the critical point where the chiral symmetry
restoration takes place, is a very interesting problem.
This problem is now actual because it is planned
to put into operation the new particle accelarators
(RHIC at Brookhaven, LHC at CERN, etc), where heavy-
ion collisions will be investigated.

This paper is organized as follows. In Section 2, we give
the Lagrangians and calculate the mixing angle. In
Section 3, we calculate the kaon polarizability using the
new value for the mixing angle. The temperature dependence
of the kaon polarizability is considered in
Section 4. We summarize and conclude in Section 5.

\section{Lagrangians and determination of the mixing angle}

For definiteness, let us consider meson vertices generated by quark
loops arising in the bosonization approach of QCD-motivated quark
models \cite{6,8}. Then, the corresponding local terms satisfy
approximate chiral symmetry of the resulting effective meson
lagrangian in the case of $m_s \neq m_u$, where $m_s$ and $m_u$ are the
constituent masses of strange or $(u,d)$-quarks, respectively. First,
let us quote only that part of the meson lagrangian which describes the
decays of scalar mesons $\epsilon(650)$, $f_0(980)$ and $a_0(980)$ into two
pions and kaons \cite{7} (see Fig.~1a)
\beq
\L{1} &=& \sum_{S=\epsilon,f_0} \,
	\left( G_\pi^S \, \vec\pi^2(x) \, S(x)
		+ G_K^S \, \bar K(x) \, K(x) \, S(x)
	\right)
\no
\\
&& + G_K^{a_0} \, \bar K(x) \, \tau_3 \, K(x) \, a_0(x)
\ ,
\label{eq1}
\eeq
where
\beq
G_\pi^\epsilon &=& 2 \, g_\pi \, Z^{1/2} \, m \, \cos\theta \ ,
\no \\
G_\pi^{f_0}    &=& 2 \, g_\pi \, Z^{1/2} \, m \, \sin\theta \ ,
\no \\
G_K^\epsilon   &=& 2 \, g_K \, Z^{1/2} \,
	\left[ (2 m - m_s) \, \cos\theta
	+ \sqrt{2} \, (2 m_s - m) \, \sin\theta \right] \ ,
\no \\
G_K^{f_0} &=& 2 \, g_K \, Z^{1/2} \,
	\left[ (2 m - m_s) \, \sin\theta
	- \sqrt{2} \, (2 m_s - m) \, \cos\theta \right] \ ,
\no \\
G_K^{a_0} &=& 2 \, g_K \, Z^{1/2} \, (2 m - m_s)
\ .
\label{eq1a}
\eeq
Besides the Goldberger-Treiman relations
$g_\pi = m/F_\pi$, $g_K= \\
(m_u+m_s)/(2F_K)$, we shall use the following parameter values
$F_\pi = 93~$MeV,
$F_K = 1.16 \, F_\pi$,
$m_u = m_d = m = 280$~MeV,
$m_s = 450$~MeV.
Here $Z$ is a renormalization coefficient arising
>from $\pi$-$A_1$-mixing ($Z\approx 1.4$ \cite{6}).

Below we shall determine the mixing angle $\theta$,
which describes the deviation of the singlet-octet mixing angle from
the ideal mixing by using recent experimental data for $f_0 \to 2\gamma$
and $f_0 \to 2 \pi$. For this, we remind that the scalar components of
the ideal mixing are related to the physical meson states through
\beq
 (\sigma_8 + \sqrt{2} \, \sigma_0)/\sqrt{3} &=&
	\epsilon \, \cos\theta + f_0 \, \sin\theta \ , \no \\
 (\sqrt{2} \, \sigma_8  - \sigma_0 )/\sqrt{3} &=&
	f_0 \, \cos\theta - \epsilon\,\sin\theta \ .
\label{eq2}
\eeq
Now, let us consider two-photon decays of scalar
mesons which are described by finite quark and meson
loop diagrams (see Fig.~1 b - d).

The amplitude of the decay $S \to 2 \gamma$
arising from quark loops takes the form
\beq
T^q{}^{\mu\nu} &=& (g^{\mu\nu} \, q_1q_2 - q_2^\mu \, q_1^\nu)
\, T^q
\label{eq4x}
\ ,
\eeq
where
\beq
T^q &=& \frac{2}{9} \,
	\frac{\alpha~a_S}{\pi F_\pi Z^{1/2}} \ .
\label{eq5x}
\eeq
Here $q_{1,2}$ are the photon momenta, $\alpha =1/137$
and the coefficient $a_S$ is given by
\beq
a_\epsilon &=& 5 \cos\theta + \sqrt{2} \, \frac{F_\pi}{F_S} \,
\sin\theta \ , \no \\
a_{f_0} &=& 5 \sin\theta - \sqrt{2} \, \frac{F_\pi}{F_S}\, \cos\theta
\ , \no \\
a_{a_0} &=& 3  \ ,
\label{eq4}
\eeq
with $F_S = 1.28 \, F_\pi$ \cite{6}.
Here we have used
low-momentum expansion around $q_1\!\cdot\!q_2 \approx 0$.
In refs., \cite{8,6,rev} it has been shown that this
prescription respects $SU(3)_F$-symmetry and leads to
reasonable results for 2 photon decays of the
pseudoscalar meson nonet.

Let us next calculate the meson loop contributions
of Figs.~1c,d which, due to
$G^S \sim g_S \sim 1/\sqrt{N_c}$, contribute in
the same order of the $1/N_c$ expansion as the
quark loop diagrams.
Taking into account the Lagrangian of electromagnetic
interactions
\beq
\L{2} &=& i \, e \, A^\mu \,
\left( \pi^- \, \partial_\mu \pi^+ -
	\pi^+ \, \partial_\mu \pi^-
	+ K^- \, \partial_\mu K^+
	- K^+ \, \partial_\mu K^- \right)
\no \\
&& + e^2 \, A_\mu^2 \, (\pi^+\pi^- + K^+ K^-) \ ,
\label{eq9x}
\eeq
one easily obtains
\beq
T^{m}{}^{\mu\nu} &=&
\left( g^{\mu\nu} \, q_1q_2 - q_2^\mu q_1^\nu \right)
\, T^m
\label{eq10x}
\ ,
\eeq
with
\beq
T^m &=&
G_M^S \, \eta \, \frac{\alpha}{2\pi} \,
\frac{2}{M_S^2} \,
\left[ x_M \, \phi(x_M) - 1 \right]
\label{eq11x}
\ ,
\eeq
where the function $\phi(x)$ is defined by
\beq
\phi(x) &=&
\left\{ \begin{array}{l}
\left[ \arctan(x-1)^{-1/2}\right]^2 \ , \quad x \geq 1 \\
\left[ \frac{i}{2} \, \ln \frac{1 - \sqrt{1-x}}{1 + \sqrt{1-x}}
	- \frac{\pi}{2} \right]^2 \ , \quad x \leq 1 \ .
\end{array}\right.
\label{eq7x}
\eeq
and $x_M = \frac{2M^2}{q_1q_2}=\left(\frac{2M}{M_S}\right)^2$,
$\eta = 2 (1)$ for pions (kaons), and $M$ is the mass of
the meson circulating in the loop.

Then we use these results for the determination
of the mixing angle $\theta$ from the decays
$f_0 \to \gamma\gamma$ and $f_0 \to \pi\pi$.
Taking into account expressions (5) and (9),
the total amplitude for the decay $f_0 \to 2 \gamma$
consisting of quark, pion and kaon loops,
reads\footnote{Small contributions of loop
diagrams with non-strange scalar mesons (10\% of
$T^\pi_{f_0\to\gamma\gamma}$) and strange
scalar mesons (4\% of $T^K_{f_0\to\gamma\gamma}$)
will be omitted.}
\beq
T_{f_0\to\gamma\gamma}
&=& T^q_{f_0\to\gamma\gamma}
+ T^\pi_{f_0\to\gamma\gamma}
+ T^K_{f_0\to\gamma\gamma}
\ ,
\no \\
T^q_{f_0\to\gamma\gamma} &=&
\frac{\alpha}{\pi F_\pi} \,
\left( -0.2 \, \cos\theta + 0.9 \,\sin\theta \right)
\ ,
\no \\
T^\pi_{f_0\to\gamma\gamma} &=&
\frac{\alpha}{\pi F_\pi}
\, ( - 0.4 + 0.19 \, i ) \, \sin\theta
\ ,
\no \\
T^K_{f_0\to\gamma\gamma} &=&
\frac{\alpha}{\pi F_\pi}
\, \left( -0.7 \, \cos\theta + 0.1 \, \sin\theta \right)
\ ,
\no \\
T_{f_0\to\gamma\gamma}
&=&
- 0.9 \,
\frac{\alpha}{\pi F_\pi}
\, \sin\theta \left( \cot\theta - 0.7 - 0.21 \, i \right)
\label{eq12x}
\ .
\eeq
Now, using the Lagrangian $\L{1}$ and the
amplitude~\gl{eq12x}, we obtain the following
expressions for the decay widths of $f_0 \to \gamma\gamma$
and $f_0 \to \pi\pi$
\beq
\Gamma_{f_0 \to \gamma\gamma} &=&
	\frac{M_{f_0}^3}{64\pi} \, \left| T_{f_0 \to \gamma\gamma}
\right|^2
\ ,
\label{eq13x}\\
\Gamma_{f_0 \to \pi\pi} &=&
\frac{3 Z}{2\pi M_{f_0}} \,
\left(\frac{m^2}{F_\pi} \right)^2 \,
\sqrt{1 - \left(\frac{2 M_\pi}{M_{f_0}}\right)^2} \, \sin^2\theta
\ .
\label{eq14x}
\eeq
By comparing their ratio with the experimental
value \cite{10}
\beq
\frac{\Gamma_{f_0 \to \gamma\gamma}}{\Gamma_{f_0 \to \pi\pi}}
&=&
1.5 \cdot 10^{-5} \ ,
\label{eq15x}
\eeq
we obtain two solutions, $\theta = 23^\circ$ and
$\theta = - 43^\circ$.
The choice $\theta = 23^\circ$ leads to predictions
$$
\Gamma_{f_0\to\gamma\gamma} =  1.1~\mbox{keV} \ ,
$$
$$
\Gamma_{f_0\to\pi\pi} = 72~\mbox{MeV} \ ,
$$
which are somewhat larger than the
averaged values of PDG \cite{10}. It is worth
mentioning that discarding meson loops would even
yield a negative angle $\theta \sim -18^\circ$.

For later applications, let us also
quote the effective Lagrangian describing the
radiative decay $K^* \to K\gamma$ of vector mesons corresponding to
the anomalous quark triangle diagram shown in Fig.~1e,
\beq
\L{3} &=&
\frac{e \, g_V}{32\,\pi^2\,F_K} \, a_{K^*} \
	\epsilon^{\sigma\kappa\nu\mu} \,
	F_{\kappa\nu}(x) \, K^*_{\sigma_\mu}(x) \, K(x) \ ,
\label{eq5}
\eeq
with $g_V=g_\rho$ being the vector meson coupling constant \\
(${g_\rho^2}/{(4\pi)} \approx 3$).
Moreover, using the value of the quark mass ratio
$\lambda = m_u/m_s \sim 0.62$, the coefficients $a_{K^*}$ are estimated
as \cite{7}
\beq
a_{{K^*}^+} &=&
\frac{1}{2} \, \left[
	1 - \frac{ 2 \, \lambda}{1 - \lambda^2} \,
	\left( 3 + \frac{2 + \lambda^2}{1-\lambda^2} \, \ln
	\lambda^2\right) \right] \simeq 1.2 \ ,
\no \\
a_{{K^*}^0} &=& 1 - \frac{\lambda}{1 - \lambda^2} \, \ln \lambda^2
	\simeq 2 \ .
\label{eq6}
\eeq

Finally, we need the box-diagram describing the
low-energy Compton-effect off $K$ mesons,
shown in Fig.~2. This diagram leads to
the Lagrangian term
\beq
\L{4} &=&
\frac{\alpha}{18 \pi \, F_K^2} \,
	F_{\mu\nu}^2(x) \, ( K^+(x) \, K^-(x)
	+ 4 \, K^0(x) \bar K^0(x) )
\label{eq7} \ .
\eeq

\section{Kaon polarizability}

In nonlinear chiral models \cite{5} (arising in the limit
$M_S \to \infty$) and in the chiral-symmetry limit of the
linear $\sigma$-model \cite{7} the main contributions to
the polarizability of charged kaons and pions emerge from pole diagrams
with intermediate scalar mesons containing the quark loop
vertices shown in
Fig.~1a,b \footnote{The contribution of meson
loop diagrams will be discussed later on (see Fig.~3).}.
This is contrary to the case of neutral mesons
where the contributions of the corresponding pole diagrams are completely
cancelled  by the box diagrams shown in Fig.~2.

Beyond the chiral limit, this situation still holds for
pions, whereas for kaons the result strongly depends on
the value of the mixing angle. We will find below that
for $\theta=23^\circ$ the scalar meson contribution
is reduced by about 50\% with respect to the value
of the chiral limit but still dominates the box diagram
contribution. For neutral kaons the
contribution of pole diagrams remains somewhat dominant
with respect to the box diagram term, leading
to a small nonvanishing value of the neutral
kaon polarizability.
To see this in more
detail, let us write the contributions of scalar meson pole and box
diagrams to the kaon polarizability $\alpha_K^{(S+b)}$ in the
form\footnote{Recall that the electric and magnetic polarizabilities
$\alpha_P, \beta_P$ of pseudoscalar mesons are obtained from the
low-energy Compton amplitude by the decomposition
\beq
T_{NR} &=&
- \vec\epsilon \cdot \vec\epsilon\,{}' \, \frac{\alpha}{M}
+ \vec\epsilon \cdot \vec\epsilon\,{}' \, \omega\omega' \,
	\alpha_P
+ (\vec\epsilon\times\vec k) \cdot ( \vec\epsilon\,{}'\times k') \,
	\beta_P
\no
\eeq
with $\omega\, (\omega')$, $k \, (k')$ and $\epsilon\,(\epsilon\,{}')$
being the incoming (outgoing) photon energy, momentum and
polarization, respectively.}
\beq
\alpha_{K^+}^{(S+box)} &=&
	C \, (2 \, \Delta_+ - 1 ) \ ,
\no \\
\alpha_{K^0}^{(S+box)} &=&
	C \, (2 \, \Delta_- - 4) \ ,
\qquad C = \frac{\alpha}{18\pi \, F_K^2 \, M_K} \ ,
\label{eq11}
\eeq
where
\beq
\Delta_\pm &=&
(m_s + m) \,
\left\{ \frac{a_\epsilon}{M_\epsilon^2}
\left[ (2m-m_s) \, \cos\theta + \sqrt{2} (2 m_s - m) \,
\sin\theta\right]
\no \right. \\
&& \left. \qquad
- \frac{a_{f_0}}{M_{f_0}^2}
\left[ \sqrt{2}(2m_s - m) \, \cos\theta - (2 m - m_s) \,
\sin\theta\right]
\no \right. \\
&& \left. \qquad
\pm \frac{a_{a_0}}{M_{a_0}^2}
\, (2m - m_s) \right\} \ .
\label{eq12}
\eeq
Taking into account \gl{eq4}, $\theta=23^\circ$ and
using two different values of the $\epsilon$-mass,
$M_\epsilon = 650~(750)$~MeV, we obtain
$\Delta_+ = 3.55~(2.6)$, \\
$\Delta_- = 3.05~(2.1)$.
This leads to the following estimates
\beq
\alpha_{K^+}^{(S+box)} &\simeq&
C \, (7.7 (5.8) - 1.1 + 0.5 - 1) =
10.5 (7.2) \cdot 10^{-4} \, \mbox{fm}^3 \ , \
\no \\
\alpha_{K^0}^{(S+box)} &\simeq&
C \, (7.7 (5.8) - 1.1 - 0.5 - 4) =
3.65 (0.34) \cdot 10^{-4} \, \mbox{fm}^3 \ .
\label{eq13}
\eeq
For illustration, let us compare \gl{eq13} with the results obtained
in the chiral symmetry limit ($\theta = 0$,
$M_\epsilon^2=M_{f_0}^2=M_{a_0}^2 \simeq 4m^2$). In this case we have
\beq
\alpha_{K^+}^{(S+box)} &\simeq&
C \, (5 + 2 + 3 - 1) = 15.4 \cdot 10^{-4} \, \mbox{fm}^3 \ , \
\no \\
\alpha_{K^0}^{(S+box)} &\simeq&
C \, (5 + 2 - 3 - 4) = 0 \ ,
\label{eq14}
\eeq
where the first three terms in the brackets of (\ref{eq13}) and
(\ref{eq14}) denote the
pole contributions of $\epsilon$, $f_0$ and $a_0$ mesons whereas the
last terms denote the contribution of the box diagrams. Thus, the
deviation from chiral symmetry in combination with the above
mixing angle influences the polarizability.
For $\theta=23^\circ$ the contribution of
the $f_0$-meson has not only decreased in absolute value
but even changed in sign compared to the chiral limit,
whereas the $a_0$-contribution is reduced by an order
of magnitude.

Scalar meson pole
diagrams and box diagrams supply also analogous contributions to the
magnetic polarizability $\beta_{K}^{(S+box)}$ but with the opposite sign
($\alpha_K^{(S+box)} = - \beta_K^{(S+box)}$).

In addition, a large contribution to the magnetic
polarizability of neutral kaons
arises from pole diagrams containing the intermediate
vector meson $K^*$. It can easily be evaluated using the Lagrangian
$\L{3}$ which leads to the result
\beq
\beta_K^{(K^*)} &=&
\left( \frac{a_{K^*}}{2\pi\,F_K}\right)^2 \,
\frac{ \alpha\,\alpha_V\,M_K}{M_{K^*}^2- M_K^2}
=
\left\{
\begin{array}{r}
4.9 \cdot 10^{-4} \, \mbox{fm}^3 \ (K^+) \\
12.7 \cdot 10^{-4} \, \mbox{fm}^3 \ (K^0) \ .
\end{array}
\right.
\label{eq15}
\eeq

For completeness, let us consider the meson loop
diagrams shown in Fig.~3,
leading to additional contributions to the kaon polarizability
comparable to the values obtained from \gl{eq11}. These contributions
have first been evaluated in ref.~\cite{4} for a nonlinear chiral
theory, providing a suitable low-energy approximation of the
linear $\sigma$-model. The expression of the
Compton-amplitude associated with the diagrams exhibited in
Fig.~3, reads
\beq
T^{\mu\nu}_{\pm}
&=&
\frac{2\alpha}{4\pi\,F^2_K} \,
\left( g^{\mu\nu} \, q_1\!\cdot\!q_2 - q_1^\nu \, q_2^\mu\right)
\, \left[ \beta^{(\pi)}(q_1q_2) + \beta_\pm^{(K)}(q_1q_2)\right]
\ , \no \\
T^{\mu\nu}_{0}
&=&
\frac{2\alpha}{4\pi\,F^2_K} \,
\left( g^{\mu\nu} \, q_1\!\cdot\!q_2 - q_1^\nu \, q_2^\mu\right)
\, \left[ \beta^{(\pi)}(q_1q_2) + \beta_0^{(K)}(q_1q_2)\right]
\ .
\label{eq16}
\eeq
The function $\beta^{(\pi)}(q_1q_2)$ arising from the above two loop
diagrams with internal pion lines is equal for charged and
neutral kaons,
\beq
\beta^{(\pi)}(q_1q_2) &=&
\frac{1}{2} \left[ \tilde x_\pi \, \phi(\tilde x_\pi) -
1\right]
\stackrel{q_1\!\cdot\!q_2 \to 0}\longrightarrow 0
\ .
\label{eq17}
\eeq
The function $\beta_\pm^{(K)}$ describes the contribution of the loop
diagrams of Fig.~3 with internal kaon propagators and
external charged kaons,
\beq
\beta_\pm^{(K)} &=&
\frac{1}{4} \, \left( \tilde x_K + 1  \right) \,
\left[ \tilde x_K \, \phi(\tilde x_K) -
1 \right]
\stackrel{q_1\!\cdot\!q_2 \to 0} \longrightarrow \frac{1}{12}
\ .
\label{eq19}
\eeq
For external neutral kaons we get
\beq
\beta_0^{(K)} &=&
\left[ \tilde x_K \, \phi(\tilde x_K) -
1\right]
\stackrel{q_1\!\cdot\!q_2 \to 0}\longrightarrow 0
\ .
\label{eq20}
\eeq
Thus, for $(q_1\!\cdot\!q_2)=0$ a nonvanishing contribution to the
kaon polarizability arises only from $\beta_\pm^{(K)}(0) = 1/12$
leading to the result
\beq
\alpha_{K^\pm}^{(K)} &=&
\frac{\alpha}{4\pi \, F_K^2 \, M_K} \,
\beta^{(K)}_\pm(0) \simeq
0.6 \cdot 10^{-4} \mbox{\ fm}^3 =
- \beta_{K^\pm}^{(K)} \ .
\label{eq21}
\eeq
For neutral kaons the meson loop contribution to the
polarizability vanishes.

We shall not calculate here the contributions of intermediate axial
vector mesons (A) to the polarizability. The corresponding pole diagrams
have been investigated in \cite{6,7},
\beq
\alpha_{K^\pm}^{(A)} &=& 0.9 \cdot 10^{-4} \mbox{\ fm}^3 \ ,
\qquad
\alpha_{K^0}^{(A)} = 0.5 \cdot 10^{-4} \mbox{\ fm}^3
\label{eq22} \ .
\eeq
We mention that a noticeable
contribution, compared to the other terms,
was obtained only for the electric polarizability of
charged kaons.
Table~\ref{t1} summarizes our estimates for the electric
($\alpha_K$) and magnetic ($\beta_K$) polarizabilities as obtained
>from quark box diagrams (box), scalar meson pole diagrams (S), meson
loop diagrams ($M$), intermediate vector (V) and axial vector (A)
mesons.
\begin{table}[htb]
\begin{center}
\begin{tabular}{||c|c|c|c|c|c|c||c||}
\hline
& box & S & M & V & A & Total & chiral limit \\
\hline
$\alpha_{K^+}$ & - 1.7 & 12.2 (8.9) & 0.6 & 0 & 0.9
& 12 (8.7) & 16.9 \\
$\alpha_{K^0}$ & - 6.8 & 10.5 (7.2) & 0 & 0 & 0.5
& 4.2 (0.9)  & 0.5 \\
$\beta_{K^+}$  & 1.7 & -12.2 (-8.9) & -0.6 & 4.9 & 0
& -6.2 (-2.9) & -11.1 \\
$\beta_{K^0}$ & 6.8 & -10.5 (-7.2) & 0 & 12.7 & 0
& 9.0 (12.3) & 12.7 \\
\hline
\end{tabular}
\end{center}
\caption{Electric and magnetic polarizabilities of kaons in
units of $10^{-4}$~fm$^3$. The columns show the contributions from
quark box diagrams (box), scalar meson poles (S), meson loops (M)
and intermediate vector (V) and axial vector (A) mesons. The pole
contributions are estimated for two values of the $\epsilon$-mass,
$M_\epsilon=650(750)$~MeV.}
\label{t1}
\end{table}
Thus, in comparision with the results of the chiral symmetry limit, for the
realistic case of physical scalar meson masses and the mixing angle
$\theta = 23^\circ$ a slightly increased value of the
electric polarizability should be observed for neutral
kaons, whereas the electric and magnetic polarizabilities
for charged kaons are reduced.
For future comparision with data let us also quote the
sum of the electric and magnetic kaon polarizabilities\
\beq
(\alpha + \beta)_{K^+} &=& 5.8 \cdot 10^{-4}~\mbox{fm}^3 \ ,
\no \\
(\alpha + \beta)_{K^0} &=& 13 \cdot 10^{-4}~\mbox{fm}^3 \ .
\eeq
Note that our results satisfy the requirement $(\alpha+\beta)_K > 0$
>from dispersion relations \cite{1}.

\section{Temperature dependence of the kaon polarizabilities}

In the previous sections we have shown that the mixing angle $\theta$
plays a very important role for the definition of the main contributions
to the kaon polarizabilities associated with the scalar pole diagrams
(see, also, \cite{Feldman}).
Unfortunately, within our model, one cannot define the temperature dependence
of the mixing angle. Let us suppose that $\theta$ has to decrease with
$T$ and to equal zero when $T = T_c$ ($T_c$ is the critical $T$).
Since the quark condensate is the order parameter in the NJL model,
also decreasing with $T$, we assume that $\theta$ is proportional
to the quark condensate or, equivalently, to the constituent quark
mass $m$: $\theta(T) = \theta~ \frac{m(T)}{m(0)} $. Clearly, this is
a very crude approximation, so that we can obtain here only qualitative
estimations of the $T$-dependence of the kaon polarizabilities.

In order to obtain the $T$-dependence of the other physical parameters
$m, m_s, M_\pi, M_K, F_\pi$ and $ F_K $ we can
use the results of our earlier
works \cite{Volkov}. The correspondiong values are given in Table II.

\begin{table}[htb]
\begin{center}
\begin{tabular}
{||l          |c        |c      |c       |c |c |c||}
 \hline
T      & $m_u$  & $m_s$   & $M_{\pi}$ & $M_K$ & $F_\pi$ & $F_K$ \\ \hline
 0   & 280    & 450    &  137   & 494   & 93         & 108    \\ \hline
 50  & 280    & 450    &  137   & 494.5 & 93         & 107    \\ \hline
100  & 271    & 448.5  &  137   & 502   & 91         & 104.6  \\ \hline
150  & 223    & 437    &  138   & 519   & 82         & 97     \\ \hline
170  & 184    & 427    &  141   & 531   & 73         & 91     \\ \hline
180  & 158    & 420.6  &  146   & 541   & 65         & 87     \\ \hline
190  & 127    & 413.5  &  155   & 555   & 55         & 82     \\ \hline
200  & 95     & 405    &  172   & 574   & 43         & 76     \\ \hline
\end{tabular}
\end{center}
\end{table}
Table II. The temperature dependence of the constituent quark
masses $m$ and $m_s$, the pion and kaon masses and their decay coupling
constants $F_\pi$ and $F_K$. All values are given in MeV.

For the scalar meson masses $M_\epsilon, M_{f_0}$ and $M_{a_0}$ we
shall use the approximate mass formulae $M_\epsilon^2 (T) = const +
4 m^2(T),~ M_{f_0}^2 (T) = const' + 4 m_s^2(T),~
M_{a_0}^2 (T) = const''+ 4 m^2(T)$, and
$M_{K^*}(T) \approx M_{K^*}(0)$.
\footnote{ As is shown in \cite{Volkov} the temperature dependence
of the scalar mesons is defined by their quark mass terms
(for instance, $M_\epsilon^2 (T) \approx M_\pi^2 (T) + 4 m^2 (T)$,
where $m(T)$ varies more rapidly wit T than $M_\pi (T)$).
The vector meson masses are stable with respect
to temperature change.} For the definition of the temperature dependence
of the axial-vector meson contributions we shall use a formula
similar to (22) (see \cite{6,7} )\
\beq
\alpha_K^{(A)} &=& C~\frac{M_K}{F_K^2 (M^2_{K_1 (1270)} - M^2_K )},\
\eeq
where $C = 3.8 ~10^{-4}$ and $M^2_{K_1 (1270)}(T) \approx
0.857~GeV^2 + 6 m(T) m_s(T)~ \\
(M_{K_1} = 1270 MeV)$.
\footnote{Here we shall ignore the weak temperature dependence
of the quark loop diagrams with the photon legs.
The temperature dependence of these diagrams is especially weak
when the relatively heavy strange quark is contained in the quark
loop. The temperature dependence of these diagrams has been
investigated in \cite{12}.}

\begin{table}[htb]
\begin{center}
\begin{tabular}
{||l          |c        |c      |c   |c |c |c |c |c||} \hline
T  & 0 & 50 & 100 & 150 & 170 & 180 & 190 & 200    \\ \hline
${\alpha}^S_{K^+}$=$-{\beta}^S_{K^+}$ & 12.2 & 12.6   & 12.8   & 12.2   & 9.8   & 6.7   & 0.54   & -9.4\\
                                      &      & (12.6) & (12.8) &  (14)  & (14.8)&  (15) & (14.6) & (13.5)    \\ \hline
${\alpha}^S_{K^0}$=$-{\beta}^S_{K^0}$   & 10.5 & 11.1   & 11.4   & 12   & 11   & 8.4   & 3.4   & -5.4\\
                                        &      &(11.2)  &(11.6)  &(14)  &(16)  &(16.8) &(17.6) &(16.5)   \\ \hline
${\alpha}^{box}_{K^+}$=$-{\beta}^{box}_{K^+}$   &-1.7& -1.7 & -1.8   & -2   & -2.2 & -2.4  & -2.6  & -3       \\ \hline
${\alpha}^{box}_{K^0}$=$-{\beta}^{box}_{K^0}$   &-6.8 &-7 &-7.2 &-8.1 &-9 &-9.6 &-10.6 &-11.9  \\ \hline
${\beta}^V_{K^+}$       &4.9 &5 &5.3 &7.2 &9.0 &10.6 &12.8 &15.8  \\ \hline
${\beta}^V_{K^0}$       &12.7 &13 &13.6 &16.4 &18.7 &20.5 &23 &26  \\ \hline
${\alpha}^A_{K^+}$      &0.9 &0.9 &1 &1.4 &1.8 &2.2 &2.3 &3.8  \\ \hline
${\alpha}^A_{K^0}$      &0.5 &0.5 &0.6 &0.8 &1 &1.2 &1.3 &2.1  \\ \hline
\hline
${\alpha}^{tot}_{K^+}$  &11.4 &11.8 &12 &11.6 &9.4 &6.5 &0.2 &-8.6  \\ \hline
${\alpha}^{tot}_{K^0}$  &4.2 &5 &4.8 &4.7 &2.9 &0 &-5.9 &-15  \\ \hline
${\beta}^{tot}_{K^+}$   &-5.6 &-5.9 &-5.7 &-3 &1.4 &6.3 &15 &28  \\ \hline
${\beta}^{tot}_{K^0}$   &9 &8.9 &9.4 &12.5 &16.8 &21.7 &30 &43  \\ \hline
\end{tabular}
\end{center}
\end{table}
Table III contains the temperature depending contributions
to the electric and magnetic kaon polarizabilities of the scalar
pole diagrams, box, vector meson pole and axial meson pole diagrams.
In the brackets we give the results corresponding to the case when
$\theta (T) = const $. The units are $10^{-4}~fm^3 $ for the
polarizabilities. Here, we have ignored the meson loop contributions.

Let us discuss the temperature dependence of different
contributions to the kaon polarizabilities. As Table III shows,
the scalar pole contributions remain approximately constant up
to $T \approx 170 MeV$ and then monotonously decrease changing
the sign at $190 MeV$ (charge kaon) and $180 MeV$ (neutral kaon).
On the other hand, the contributions of the box diagrams,
the vector and axial vector pole diagrams monotonically
increase in their absolute values. In summary, the total
electric kaon polarizabilities are approximately stable in the
temperature interval $0 < T < 150 MeV$ and then monotonically
decrease, changing their signs for $T > 180 MeV$
(charge kaon) and $T > 170 MeV$ (neutral kaon). The magnetic
kaon polarizabilities monotonically increase in the whole temperature
domain. In particular, the charged magnetic polarizability
changes the sign for $T > 150 MeV$. Note that for a constant mixing
angle (see values in brackets) the scalar meson pole contributions
would weakly increase until $T \approx 190 MeV$.

\section{Summary and Conclusions}

In this paper, we have estimated the mixing angle of scalar mesons
within a chiral quark $\sigma$-model
>from the decays $f_0 \to \gamma\gamma$,
$f_0 \to \pi\pi$ using new data \cite{10} and taking into
account both quark and meson loop diagrams.
The new value of the mixing angle $\theta = 23^\circ$ obtained
by including meson loops, significantly differs from the value
$\theta = - 18^\circ$ which would have been obtained if only
quark loops were taken into account. The value $\theta = 23^\circ$
is then used to estimate the polarizability of kaons. For
this aim, we have calculated the low-energy Compton amplitude
for $K$ mesons, taking into account scalar pole diagrams as well
as box and triangle diagrams including quark and meson loops.
As Table~\ref{t1} shows, scalar meson poles give an important
contribution to the electric and magnetic polarizabilities of
charged and neutral mesons. The dominant scalar meson
contributions arise here from $\epsilon$ and $f_0$, whereas
the contribution of $a_0$ is relatively small.

Notice that our model allows one to study the deviation
>from the chiral symmetry limit ($M_\epsilon=M_{f_0}=M_{a_0}=2m$;
$\theta=0$). As we have found, for the realistic case of physical
scalar meson masses and a mixing angle $\theta = 23^\circ$
the absolute values of the electric and magnetic polarizabilities
for charged kaons are reduced by about 50\%, whereas the
electric polarizability of neutral kaons increases.
In the latter case, the results turn out to be rather sensitive
to changes of the $\epsilon$-mass.

It is interesting to compare our results with predictions of other
models. For example, the results of the Quark Confinement Model (QCM)
\cite{11} for the charged pion polarizability are close to
the predictions of our model in ref.~\cite{7,12}, whereas
they are larger by a factor 2--4 in the case of the kaon polarizability.
An analogous result as in \cite{11}
was obtained for neutral kaons in an $SU(3)$ NJL-model
\cite{13}.

In conclusion, let us emphasize that we have obtained our results
in the framework of the standard NJL model. In this model one considers
only the first terms of the momentum expansion of the quark loop
diagrams without taking into account their dependence on the
momentum of the external meson legs. However, it is worth noticing
that taking into account such a momentum dependence for box and
triangle quark diagrams could lead to an additional
momentum-dependent factor $f(p^2_K)$  (see \cite{12})\
\beq
f(p^2_K) &=& 6 (m + m_s) \int_0^{\infty} dk k^2 \frac{E_u + E_s}
{E_u E_s [(E_u + E_s)^2 - p^2_K]^2}\ ,
\eeq
where $E_i = \sqrt{m_i^2 + k^2}$ and $p_K$ is the kaon momentum.

Then, our formulae (18) and (22) take the form\
\beq
\alpha^{S+box}_{K^+} &=& C~(2~\Delta_+ - f(M^2_K))\
\eeq
\beq
\alpha^{S+box}_{K^0} &=& C~(2~\Delta_- - 4~f(M^2_K))\
\eeq
and\
\beq
\alpha^{K^*}_K &=& (\frac{a_{K^*}}{2 \pi F_K})^2~
\frac{\alpha \alpha_V M_K}{M^2_{K^*} - M^2_K}~(f(M_K^2))^2\
\eeq
Neglecting the $p^2_K$ dependence in $f(p^2_K)$ reproduces our
former results (18) and (22) because $f(0) \approx 1$. However,
the function $f(M^2_K)$ has a nontrivial dependence on the
temperature. Indeed, the function $f(M^2_K)$ has a singular
behaviour near the so-called Mott-temperature when $T > T_{Mott}$,
where $M_K(T) > m_u(T) + m_s(T)$. ( The Mott temperature is the
temperature, when the kaon mass becomes equal to the sum of masses of its
quark components: $M_K(T_{Mott}) = m_u(T_{Mott}) + m_s(T_{Mott})$.
In our model $T_{Mott} \approx 187 MeV $.)
Then, the contributions to the kaon polarizabilities of the
corresponding diagrams could very strongly increase near the
Mott-point (see Table IV, where values of the function $f(M^2_K)$ are
given at different values of $T$).

\begin{table}[htb]
\begin{center}
\begin{tabular}
{||l          |c   |c    |c      |c   |c      |c |c |c||} 	\hline
 T       & 0   & 50   & 100 & 150  & 170  & 180  & 185  & 190  \\ \hline
$f(M^2_K)$ & 1.6 & 1.62 & 1.65& 2.03 & 2.74 & 4.21 & 8.3  & $\infty$ \\ \hline
\end{tabular}
\end{center}
\end{table}
Table IV. Values of the function $f(M^2_K)$ at different $T$.

 The physical meaning of this effect
is transparent. Before the transition into the quark-gluon plasma
the dipole momentum of the meson strongly increases, and then we
observe the dissociation of the meson into their quark components.

 Clearly, a more quantitative discussion of the temperature
behaviour of the kaon polarizabilities near the Mott point is
beyond the scope of the local NJL model. It would be interesting
to investigate this effect more carefully in realistic nonlocal
quark models including confinement and meson form factors.


 One of the authors (M.K.V.) would like to thank Prof. J. H\"ufner
for the useful discussions. This work was
partially supported by the European Community via INTAS (94-2915).

\newpage
 Fig.1 The triangle quark-loop diagrams, describing the decays:\\
a) $S \rightarrow 2 \pi~(2 K)$, b) $S \rightarrow 2 \gamma$,
e) $K^* \rightarrow \gamma K$. \\
c) The triangle meson-loop diagrams,
describing the decays $S \rightarrow 2 \gamma$.  \\
d) The meson-loop diagrams,
describing the decays $S \rightarrow 2 \gamma$. \\

 Fig.2 The quark-box diagrams, describing the Compton effect off kaons.\\

 Fig.3 The meson-loop diagrams, describing the Compton effect off kaons.

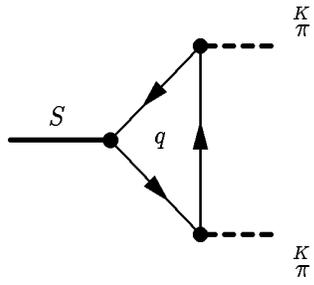
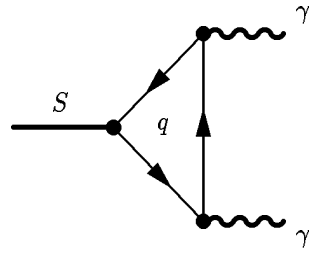
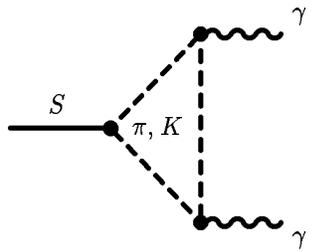
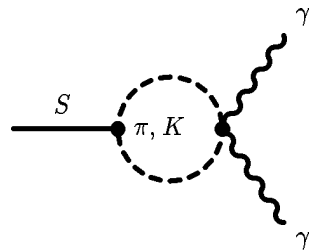
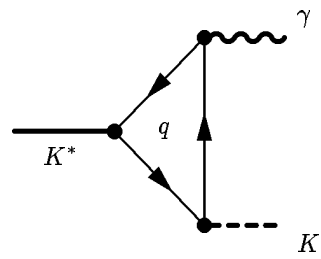
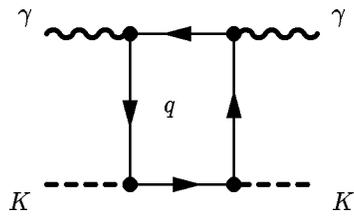
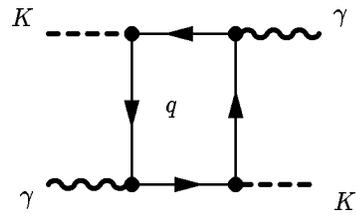
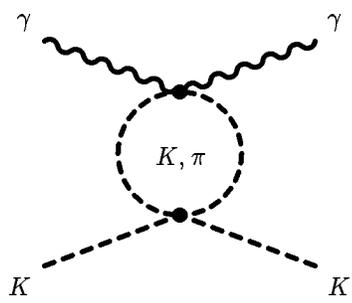
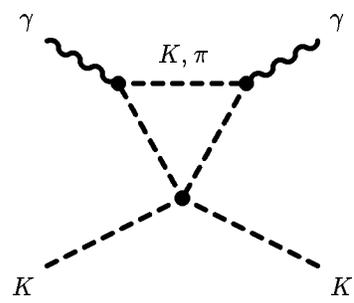
\end{document}